\documentclass[sigconf]{acmart}

\def\BibTeX{{\rm B\kern-.05em{\sc i\kern-.025em b}\kern-.08emT\kern-.1667em\lower.7ex\hbox{E}\kern-.125emX}}
    
\usepackage{booktabs} 
\usepackage{graphicx}
\usepackage{multirow}
\usepackage{siunitx}
\usepackage{subcaption}
\usepackage[most]{tcolorbox}
\usepackage[export]{adjustbox}
\usepackage{amssymb}

\setcopyright{rightsretained}

\acmConference[ ]{   }{ }{ }
\acmYear{2019}
\copyrightyear{ }

\acmPrice{15.00}

\begin{document}
\title{Towards an Automated Unified Framework to Run Applications for Combinatorial Interaction Testing}

%
\author{Bestoun S. Ahmed*}

\orcid{1234-5678-9012}
\affiliation{%
  \institution{Dept of Mathematics and Computer Science, Karlstad Univ, Sweden and
  Dept of Computer Science, Czech Technical Univ in Prague, Czech Republic}
}
\email{bestoun@kau.se}

\author{Amador Pahim}
\affiliation{%
  \institution{Red Hat Czech s.r.o., Brno, Czech Republic}
}
\email{apahim@redhat.com}

\author{Cleber R. Rosa Junior}
\affiliation{%
  \institution{Red Hat, Inc., Westford, USA}
  }
\email{crosa@redhat.com}

\author{D. Richard Kuhn}
\affiliation{
  \institution{Natl Inst of Standards and Technology, Gaithersburg, MD, USA }
  }
\email{kuhn@nist.gov}

\author{Miroslav Bures}
\affiliation{%
  \institution{Dept of Computer Science, Faculty of Electrical Eng, Czech Technical Univ, Prague, Czech Republic}
}
\email{buresm3@fel.cvut.cz}

\renewcommand{\shortauthors}{B. Ahmed et al.}

\begin{abstract}
Combinatorial interaction testing (CIT) is a well-known technique, but the industrial experience is needed to determine its effectiveness in different application domains. We present a case study introducing a unified framework for generating, executing and verifying CIT test suites, based on the open-source Avocado test framework. In addition, we present a new industrial case study to demonstrate the effectiveness of the framework. This evaluation showed that the new framework can generate, execute, and verify effective combinatorial interaction test suites for detecting configuration failures (invalid configurations) in a virtualization system. 
\end{abstract}

%
%

%

\maketitle

\section{Introduction}
Combinatorial interaction testing (CIT) (sometimes called $t$-way testing)  is based on the \textit{covering array} (CA) \cite{HARTMAN2004149}, a matrix that includes all $t$-way combinations of input parameter values, for a specified level of $t$ (usually $t \leq 6$ for software testing) of the system-under-test (SUT). Research activities have focused mainly in two directions (1) generating combinatorial interaction test suites that provide $t$-way coverage, and (2) applying a combinatorial interaction approach to test industrial systems.

CIT has shown impressive results in many testing studies and large-scale industrial projects \cite{bartholomew2013industry,li2016applying}. The usefulness of CIT could be for example a systematic reduction of a test suite or detecting new faults in a SUT due to interactions of input parameters, or identifying invalid configurations of the SUT. In fact, finding new applications for CIT is an active research area. 

To apply CIT on any SUT, the first step is to model the input parameters or configurations of the system. Typically, this process involves identifying the input parameter and configuration values that are needed in testing.  For continuous-valued parameters, equivalence class partitioning will generally be needed to reduce the domain size to a tractable level.  The CA generation tool uses the input model to produce test suites that cover all $t$-way combinations of values of the CA variables. A limited form of this method is "pairwise" or "all-pairs" testing, which includes all 2-way combinations of parameter values. Stronger forms of combinatorial testing use 3-way, 4-way, or higher strength CAs to detect complex faults that depend on multiple factors interacting.  

Once the test suites are generated, they must be executed and their output verified, steps which can be combined and automated. For example, test generation and execution can be combined using a scripting language (e.g.,\cite{Anwar2016}). However, applying these steps may vary from one application to another, so implementing CIT in practice is usually application-specific.

In this paper, we introduce a generic unified framework approach to apply CIT in practice. This framework is an output of a successful industry-academia collaboration effort. We have integrated the CIT capabilities into the Avocado\footnote{https://avocado-framework.github.io/} framework. Avocado is an open source testing framework maintained by Red Hat Inc. and Avocado Community Contributors. The framework consists of a combination of tools and libraries to ease automated testing by providing a set of programs for test execution and different APIs for writing test cases. The test can be written in a user's programming language or using a Python API. The new plugin will add the capability of CIT to the framework. Originally, Avocado was designed as a flexible framework that can be used to run any set of test cases for an application as far as the plugin of that application is available. For example, the user can use it for unit testing, virtualization testing, security testing, or mobile application testing. Almost everything is a plugin in Avocado and the development of a new plugin is made straightforward to extend the functionality of the runner for testing any new application. 

Using these capabilities of Avocado, we have designed and implemented a new and unique plugin to extend the Avocado framework by interacting with the other plugins, which allows generating much more efficient and effective test data. In doing so, the user will need only to follow the input modeling style of Avocado to enter the values of the SUT and then let Avocado generate, run, and verify the test cases.

\section{Background and literature}\label{Background}

CIT relies on a covering array (CA), to derive the combinatorial interaction test suites. A $CA$ is based on $t$-way coverage criteria (where $t$ represents the desired interaction strength, which is the number of factors interacting). $CA (N; t, k, v)$, also expressed as $CA (N; t, v^k)$, is a combinatorial structure constructed as an array of $N$ rows and $k$ columns on $v$ values such that every $N \times t$ sub-array contains all ordered subsets from the $v$ values of size $t$ at least once. A mixed-level covering array $(MCA) (N; t, k, (v_1, v_2,\ldots \ v_k))$ or $MCA (N; t, k, v^k)$ may be adopted when the number of component values varies, while the Constrained Covering Array (CCA) may be adopted when there are constraints among the values of input parameters \cite{BestouIEEAcess}. 

Research activities in CIT include (1) the generation of combinatorial interaction test suites, and (2) the application of CIT \cite{BestouIEEAcess}. In fact, the problem of test suite generation has received more attention from the research community. 
Generation algorithms vary from random generation \cite{Huang2012} to mathematical constructions for limited small size and interaction strength such as generation from  Orthogonal Arrays (OA) \cite{HARTMAN2004149}, to the deterministic generation methods like In parameter Order (IPO) algorithm and its variants IPOG, IPOG-D \cite{Lei:2008:IET}, and IPOG-F \cite{Forbes2008287}. Metaheuristic algorithms have also been used widely in the last decade to optimize the generation process. Here, many algorithms are used, such as Genetic Algorithms (GA) \cite{Bansal2014}, Ant Colony Algorithms (ACA) \cite{Shiba:2004}, Particle Swarm Optimization (PSO) \cite{Ahmed:2011}, and Tabu Search \cite{NURMELA2004143}. A comprehensive survey of these generation algorithms and many others can be found in \cite{BestouIEEAcess}. In fact, with the availability of all these algorithms and tools, algorithm research for CIT can be said to have reached a mature state, making CIT practical for real-world applications. 

Many studies have investigated different applications of combinatorial methods to software testing and program verification. Many applications emerged in this direction, including investigation of the relationship between code coverage and $t$-way coverage \cite{Huller00reducingtime}, fault detection and characterization \cite{Yilmaz:2004}, graphical user interface testing (GUI) \cite{Yuan:2011}, and model-based testing and mutation testing \cite{Bures2017}. In fact, there are many applications of CIT in the software testing discipline. Combinatorial interaction testing also finds its way to other fields rather than software testing. For example, it has been used in satellite communication testing, hardware testing \cite{Borodai1992}, advanced material testing \cite{Schubert2004}, and dynamic voltage scaling (DVS) optimization \cite{Sulaiman2013}. Many other application domains can be found in the literature, and researchers are actively discovering new uses for CIT. 

As mentioned previously, to apply CIT in practice, there is a need to organize variables and values of the SUT into some interpretable input model for input to the CA test generation tool. After generating the test cases, tests should be executed on the SUT, and then the test output should be verified for the pass and fail criteria. These steps are applicable for almost all applications. However, details of each step may vary from one application to another. For example, the input model, and the execution of the test cases may vary depending on the SUT.  There are always some manual activities in this process. In the literature, there are some efforts \cite{Huang2012} to develop adaptive solutions to generate and execute the test cases, but still, they are application-specific solutions. A more practical solution is to develop a flexible automated framework that can generate, run, and evaluate combinatorial test suites on any SUT. One approach to integrating these steps is CITLab \cite{gargantini2012citlab}, designed to improve the interoperability among combinatorial testing tools, by providing a framework for defining domain-specific languages. 

In this paper, we introduce an enhancement of the Avocado testing framework to include CIT capability and apply this new capability to automate testing of configuration specifications for the open source hypervisor Qemu. Avocado provides a set of tools and supporting libraries for test automation on the Linux platform. The framework can take the input of the SUT as a model and generate test cases according to it, then run and verify the test cases, as described in the following sections.

\section{The Avocado Framework}\label{AvocadoFramework}

\begin{figure}
\centering
\includegraphics[width= 2 in]					{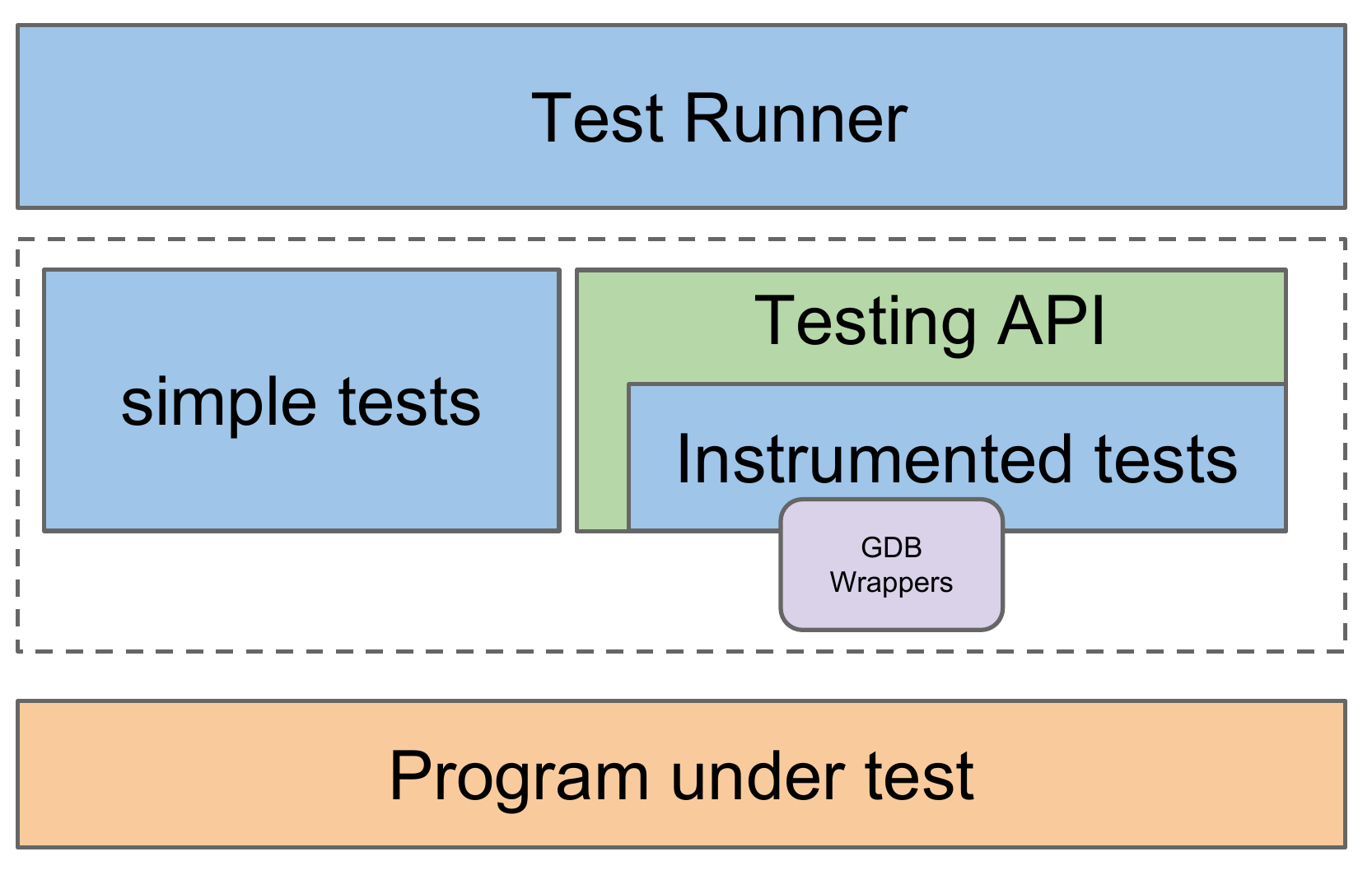}
\caption{Avocado high level test writer view}
\label{Figure:AvocadoStructure}
\end{figure}

Avocado is an open source testing framework maintained by Red Hat Inc. and Avocado Community Contributors, that is designed to give common ground to both quality assurance (QA) teams and Developers. The framework is a set of tools and libraries to help with automated testing. Here, the native tests are written in Python; however, any executable can serve as a test. Avocado consists of three main components, the test runner, libraries, and plugins. The test runner enables the user to run the test cases. Avocado provides the flexibility of writing test cases in Python or any other programming language. In both cases, there are facilities to record the activities during the test, such as information collection of the system and automatic logging. Libraries are used to create and write test cases in an expressive way. The plugins are extensions to Avocado for adding more functionality and features to the framework. The ability to add more plugins to the framework easily assures maximum flexibility for future developers. Figure \ref{Figure:AvocadoStructure} shows the basic building blocks of the framework from the user perspective. The framework supported by the GDB\footnote{https://sourceware.org/gdb/} front-end for the user interface. The test runner relies on plugins and many of them can be used during test execution. The output of the testing process can be saved and used in JSON, Xunit, HTML, or TAP formats. Hence, the output file can be used in different ways by the developers or testers. For example, it can be integrated with Jenkins\footnote{https://jenkins.io/} to trigger the testing process.

To automate the testing process of any SUT, the tester needs to create test cases and then run them using the plugins. The plugin for that SUT must be used to automate the process. The test runner will automatically run the SUT with the necessary environment (i.e., necessary resources and programs) to execute the test cases. For example, suppose we want to test a mobile application using some set of test cases and record the output in any of the used formats. Here, the test runner will run the test cases using a plugin to start the mobile emulator. When the test runner initiates the emulator, the test cases will be executed automatically by the Avocado, and the output will be saved directly to the user output format. 

If the user of Avocado wants to test an application and the plugin is not available, then the user first needs to create a plugin to start the environment and setting up all the relevant services, stubs, and programs to start the application. In fact, the creation of a new plugin for this purpose is simplified by our development team. The user needs to follow a few simple steps to create this plugin. These steps can be found in the Avocado documentation\footnote{http://avocado-framework.readthedocs.io/en/59.0/}. 

These flexibilities of the Avocado framework make it attractive to implement an automated unified framework for combinatorial interaction testing applications. Here, we extend Avocado to handle the combinatorial interaction testing automatically to perform the model creation, test case generation, test case execution, and evaluation of the test oracle. By adding this feature, we can get the benefits of CIT to reduce and detect faults in the SUT, with the benefits of Avocado to automate, run and verify the testing process. In the next section, we illustrate this CIT extension to Avocado.

\section{The CIT Extension to Avocado}\label{CITExtentionOfAvocado}

\begin{figure}
\centering
\includegraphics[width= 2.8 in]					{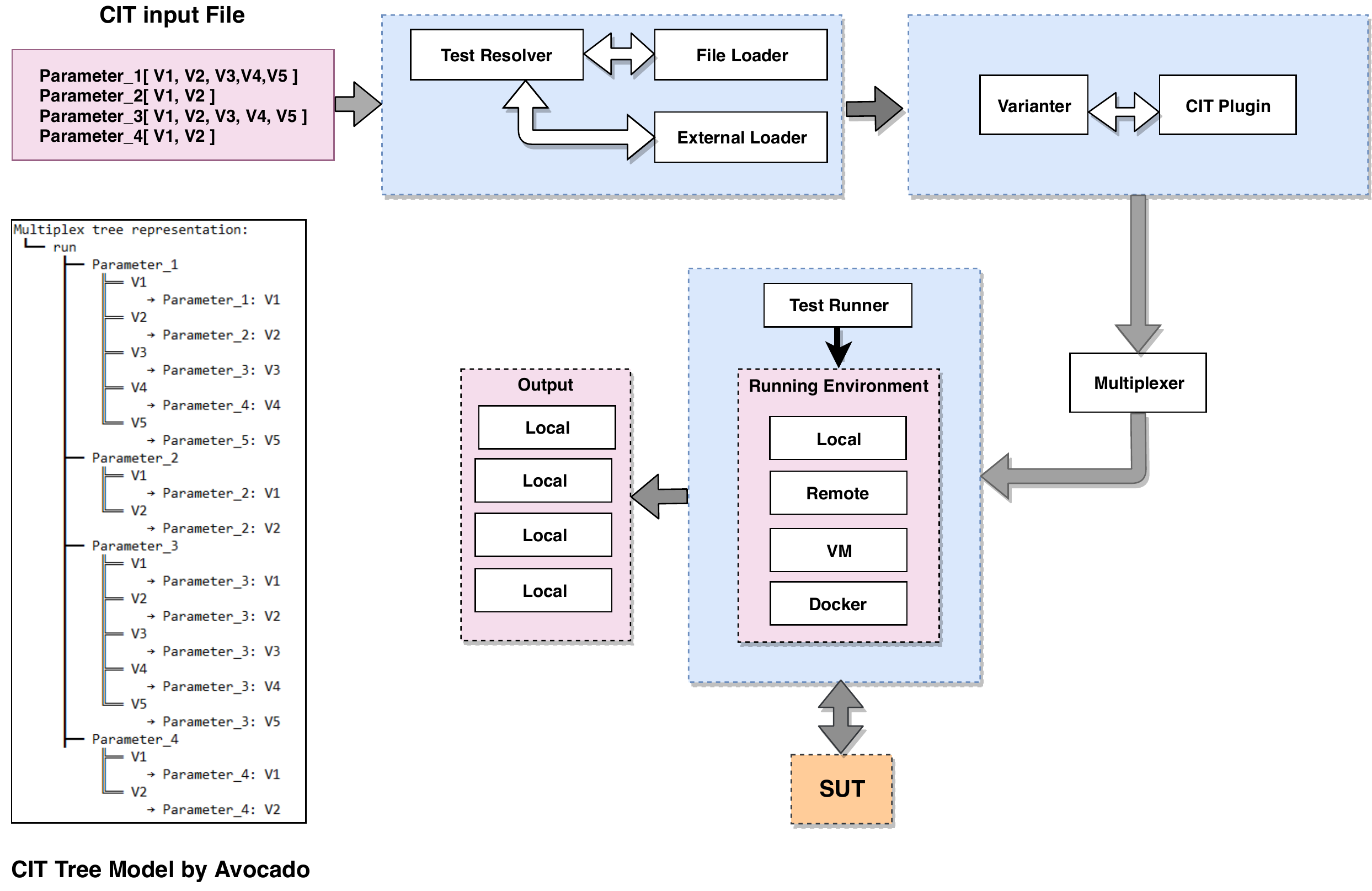}
\caption{Basic structure of Avocado with CIT extension }
\label{Figure:CITWithAvocado}
\end{figure}

The CIT extension to Avocado adds significant capabilities to the framework and makes it possible to apply CIT in a fully automated testing process (see Figure \ref{Figure:CITWithAvocado}).
Here, the user first needs to model the SUT input parameters to the CIT file format, i.e., determine the parameters to be included in tests, and values for each parameter. The output of CIT test generation will be a set of tests that covers all $t$-way combinations of the parameter values, for a selected level of $t$.  Thus, if $t$=2, the tests will instantiate all sets of variables taken two at a time with all pairs of possible values for these variables.  For example, in Figure \ref{Figure:CITWithAvocado}, there are four input parameters, each with a different number of values. The input parameters and values are also represented in Avocado in a tree representation model. The test resolver will resolve this input model to an interpreted format understandable by the CIT plugin. The CIT plugin will generate the test cases and send them to the multiplexer that creates the scripts to be run on the SUT. In some applications, the test suite is a configuration setting of the SUT and there are still some input files or codes that must be input with the configuration. We call this type of input a \textit{variant}. Hence, if there is a variant in the SUT, Avocado will consider it with the test suite through the multiplexer. The multiplexer will then send the test cases to the test runner. The test runner executes the test cases with respect to the SUT. For each test case, the Avocado will record the output and show a verification message on the screen. 

The test runner can run the test cases in different ways. As shown in Figure \ref{Figure:CITWithAvocado}, it can run the test cases on the conventional local machine, remote machine, virtual machine, or even on a Docker container. After running the test cases, the runner saves the recorded results of the output in a customizable format. The output format can be in XML, JSON, Tap, or HTML format, as selected by the user. The final runner output and test time execution are also shown on On the output screen in Pass and Fail forms. Here, the PASS and FAIL depend on the output of the test case. For a simple test, it is PASS when the exit code is "0". For an instrumented test, it is considered PASS for example when there is no exception in setUp(), testMethod(), or tearDown(). A test is also considered PASS when the runner gets the final test status. This PASS and FAIL is also customizable, and the user can state the PASS and FAIL based on the output summary of the test.

Combinatorial interaction test suites are generated with the previously developed algorithm PSTG \cite{Ahmed:2010:PTS,Ahmed:2012}. PSTG generates combinatorial interaction test suites using Particle Swarm Optimization (PSO). The algorithm has been assessed extensively and proved its efficiency, with different benchmarks and experiments in the literature. The user also can contribute to the test generation algorithm as it is an open source framework.

\section{An Industrial Case Study} \label{CaseStudies} 

To show the effectiveness of the automated framework, it is applied in an industrial case study of validating configurations in a virtualization environment. In addition to the evaluation aim of this case study, it also demonstrates a somewhat different application of CIT, in which CAs are used in configuration checking as compared with conventional code testing.

\subsection{Object of the Study} \label{subsec:object}

The object of the study is the Qemu virtualization Project\footnote{https://www.qemu.org/}, specifically, the Qemu Block Layer configuration checking tests. Qemu is a machine emulator and virtualizer. An essential component of this virtualizer is its Block Layer. Every emulated or virtualized machine will need at least one virtual disk to fulfill its purpose. Qemu supports a variety of disk image formats, and they can be created in many different ways.  To ensure that a configuration can function correctly on the virtualizer, a set of shell scripts and commands are run on Qemu. In this case study, there are 192 such validation scripts. A configuration is determined to be \textit{valid} if all 192 validation scripts pass, otherwise the configuration is \textit{invalid}.

We present a real case study from the industry. In this work, CIT tests are not used to detect coding errors, but to identify invalid configuration settings. The case study is a sample of bigger projects that were inspired and adopted by the Red Hat Quality Assurance (QA) team. We used the Avocado framework as a test bed.

To cover with the required testing of those many formats and options for creating Qemu disk images, the Qemu Project holds a `qemu-iotests` directory, with the validation test cases\footnote{The used scripts can be found in  https://github.com/qemu/qemu/tree/master/tests/qemu-iotests}. The tests consist of BASH scripts that will be executed by the Avocado Test Runner. Avocado will receive the combinations from the CIT Plugin, parse it into a Tree object and iterate that Tree object to create the Test Suite, an object containing each variation of tests per parameters combination. Test scripts will then be executed with a given `Variant`(i.e., a combination of parameters and variants) in place, to be consumed in the form of environment variables. The test script will create the virtual disk image following those options and then manipulate it to test if the generated image file complies with the requirements, as specified within the test script. Based on the test assessment of the created image manipulation, the test script will return the corresponding exit code to Avocado. Using that exit code, Avocado will mark the final test status as PASS (exit code 0) or FAIL (exit code non-0).

\subsection{The System Under Test}

The virtualized system under test used for this experiment consists of five input parameters. Each parameter represents a specific type of image that can be used as an option in the system configuration that has to be run as a Qemu project. The Avocado tree representation\footnote{Example of the tree model can be found here https://bit.ly/2UmWrEW} of the input model can be generated by Avocado.

As can be seen in the tree representation model, the system has five inputs, each of them having different configuration values. For example, the image format could be one of five values, i.e., raw, qcow, qcow2, luks, and vmdk. The full meaning of each parameter and its corresponding values is shown in Table \ref{ParameterValueMeanning}. A possible system configuration is a combination of these variables and settings, so the input model structure for combinatorial test generation would be designated $5^2 2^3 $, i.e., two variables with five values and three with two values. Note that this expression also gives the number of possible combinations, in this case 200. Combining these variables together will form a configuration; however, this does not mean that the configuration is a valid configuration, thus the need for validation scripts as explained above.

\begin{table*}
\centering
\caption{Parameter and corresponding value meanings of the SUT}
\scriptsize
\begin{tabular}{|l|>{\raggedright}p{11cm}|}
\hline 
\textbf{Parameter/Value} & \textbf{Meaning}\tabularnewline
\hline 
\textbf{img\_format} & The format in which the Qemu image will be created.\tabularnewline
\hline 
raw & no specific format, raw data.\tabularnewline
\hline 
qcow & The versatile Qemu Copy On Write format, first version.\tabularnewline
\hline 
qcow2 & The versatile Qemu Copy On Write format, second version.\tabularnewline
\hline 
luks & The Linux Unified Key Setup encrypted image format.\tabularnewline
\hline 
vmdk & VMWare image format.\tabularnewline
\hline 
\textbf{img\_protocol} & The protocol used to access the image.\tabularnewline
\hline 
file & Direct access through the filesystem.\tabularnewline
\hline 
nbd & Network Block Device protocol, enabling a remote server to be used
as block device.\tabularnewline
\hline 
\textbf{cache\_mode } & The method used to cache data with the image file.\tabularnewline
\hline 
none & The host page cache is skipped and writes happen directly from the
userspace buffers and the image.\tabularnewline
\hline 
writeback & The default option, where direct cache and no-flush are off. The host
page cache is used and writes are reported to the guest as complete
when they are committed to the page cache.\tabularnewline
\hline 
writethrough & Doesn't batch metadata updates. Writes are reported as complete after
data is committed to the image.\tabularnewline
\hline 
directsync & Writes reported as complete when the data is committed to the image,
skipping the host page cache.\tabularnewline
\hline 
unsafe & Same as "writeback" with additional ignore
for the flush commands (no-flush enabled).\tabularnewline
\hline 
\textbf{misalign} & Misalign allocations for direct writes.\tabularnewline
\hline 
true & Enabled\tabularnewline
\hline 
false & Disabled\tabularnewline
\hline 
\textbf{qemu\_img} & The qemu-img binary to create images with.\tabularnewline
\hline 
/usr/bin/qemu-img  & The Fedora 27 version (2.10).\tabularnewline
\hline 
/git/qemu/qemu-img  & The latest upstream version (master).\tabularnewline
\hline 
\end{tabular}
\label{ParameterValueMeanning}
\end{table*}

\subsection{Evaluation and Analysis}

To evaluate the effectiveness of Avocado for CIT, we present here the results of our case study. As previously mentioned, we are not evaluating the test generation algorithm efficiency of the Avocado framework, as it has been reviewed extensively in the literature. We used the PSTG algorithm to generate the test cases. More evaluation results and comparison with other algorithms and tools can be found in the literature (e.g., \cite{Ahmed:2010:PTS,Ahmed:2012}). Note also that the Avocado framework is composed of many plugins and tools, and evaluating them is beyond the scope of this paper. 

Here, we aim to validate the effectiveness and performance of the Avocado CIT extension regarding four critical issues:

\begin{enumerate}
\item The coverage validity of the generated test cases 
\item The performance of the testing process
\item The efficiency of the multiplexer integration with test generation when generating different $t$-way test suites.
\item Level of fault detection for different $t$-way test suites.
\end{enumerate}

As the generation algorithm uses PSO concepts, it generates nondeterministic results due to the random initialization of the algorithm. To this end, we ran each test several times and addressed the results to assure a fair statistical experiment. For coverage validity, we checked it in each run; however, due to limited space in the paper, we only present one graph for each $t$-way test suite. We ran the test cases for the performance and effectiveness 30 times and then produced a box-plot for them. All the experiments were performed on a Linux Fedora personal computer with 2.9 GHz Intel Core i5 CPU and 8GB 2133 MHz of memory.

\begin{figure*}[tbph]
	\centering
	\begin{subfigure}[b]{0.3\textwidth}
		\includegraphics[width=\textwidth]					{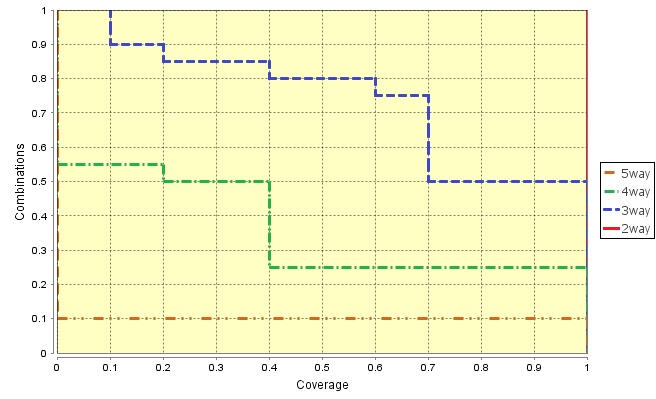}
                \vspace{-0.5cm}
		\caption{2-way full coverage \centering }
		\label{2wayCoverage}
    \end{subfigure}%
         \hfill
    \begin{subfigure}[b]{0.3\textwidth}
    	\includegraphics[width=\textwidth]					{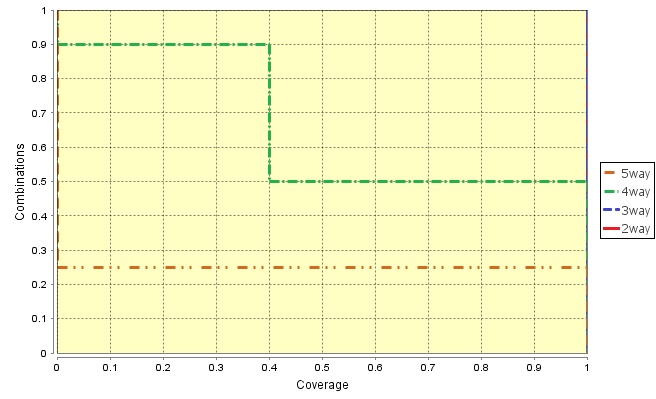}
                \vspace{-0.5cm}
        \caption{3-way full coverage \centering }
        \label{3wayCoverage}
          \end{subfigure}
             \hfill
    \begin{subfigure}[b]{0.3\textwidth}
    	\includegraphics[width=\textwidth]					{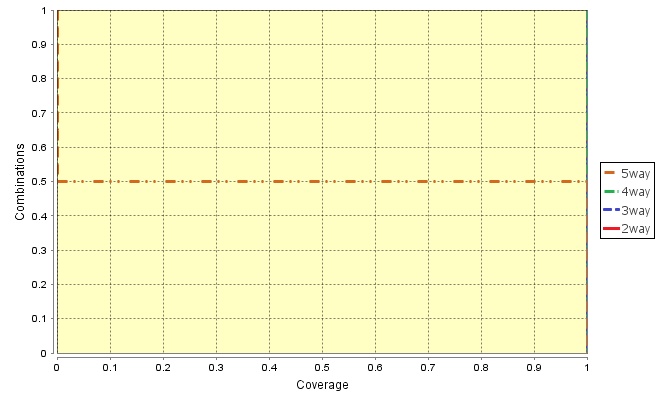}
                \vspace{-0.5cm}
        \caption{4-way full coverage \centering }
        \label{4wayCoverage}
    \end{subfigure}
\caption{The coverage measure for 2-way, 3-way, and 4-way test suites}
\label{CoverageMeasureByNIST}
\end{figure*}

The basic concept of the CIT is that all the combination tuples must be covered by the generated test suite at least once. To assure that our Avocado framework covers all these tuples, we used the Combinatorial Coverage Measurement Command Line Tool (CCMCL)\footnote{https://github.com/usnistgov/combinatorial-testing-tools} for evaluation. The tool was developed by the National Institute of Standards and Technology (NIST) to measure and validate the coverage of a $t$-way test suites. Figure \ref{CoverageMeasureByNIST} shows the coverage measure for 2-way, 3-way, and 4-way test suites.

The CCMCL tool is a coverage strength meter to determine the combinatorial coverage of any test suite and also identify any missing combinatorial tuples. Note that combinatorial coverage as evaluated by CCMCL is different from conventional structural coverage measures such as statement or branch coverage.  Combinatorial coverage is a measurement of the proportion of $t$-way combinations included in a test suite (a static measure), rather than a dynamic execution-related code measure such as statement coverage. It is clear from Figure \ref{2wayCoverage} that the generated 2-way test suite achieves the 100\% coverage of the tuples. Here, the red indicator line is on the right side of the graph which indicates the 100\% achieved coverage level. Figure \ref{2wayCoverage} also shows the 3-way, 4-way, and 5-way (i.e., the exhaustive test suite with full strength) for the same 2-way test suite for comparison. Here, we can see that the 2-way test suite can assure the full coverage of 2-way combinatorial tuples but it only covers 50\% of 3-way combinatorial tuples (blue line), 25\% of 4-way combinatorial tuples (green line), and 10\% of 5-way combinatorial tuples (brown line). The full coverage of 3-way and 4-way combinatorial tuples can be achieved with 3-way and 4-way test suites respectively in Figures \ref{3wayCoverage} and \ref{4wayCoverage}. Figure \ref{4wayCoverage} also illustrates a basic property of CAs. A CA for \textit{t}-way combinations will also provide 100\% coverage of \textit{s}-way combinations, for any $s<t$, i.e., the designated strength and all lower strength tuples. For example, in Figure \ref{4wayCoverage} the 4-way test suite covers the 4-way, 3-way and 2-way combinations. 

\begin{figure*}[tbph]
	\centering
	\begin{subfigure}[b]{0.3\textwidth}
		\fbox{\includegraphics[width=\textwidth]					{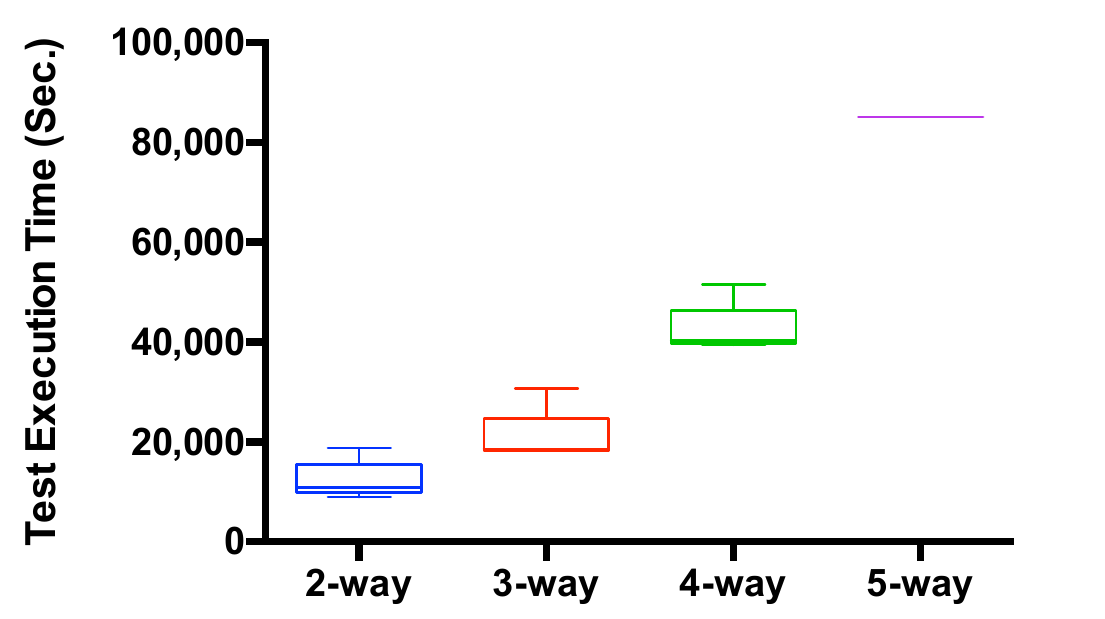}}
                \vspace{-0.15cm}
		\caption{Comparison of the test execution time by each  $t$-way test suite \centering }
		\label{Figure:ExecutionTimeBox}
    \end{subfigure}%
         \hfill
    \begin{subfigure}[b]{0.3\textwidth}
    	\fbox{\includegraphics[width=\textwidth]					{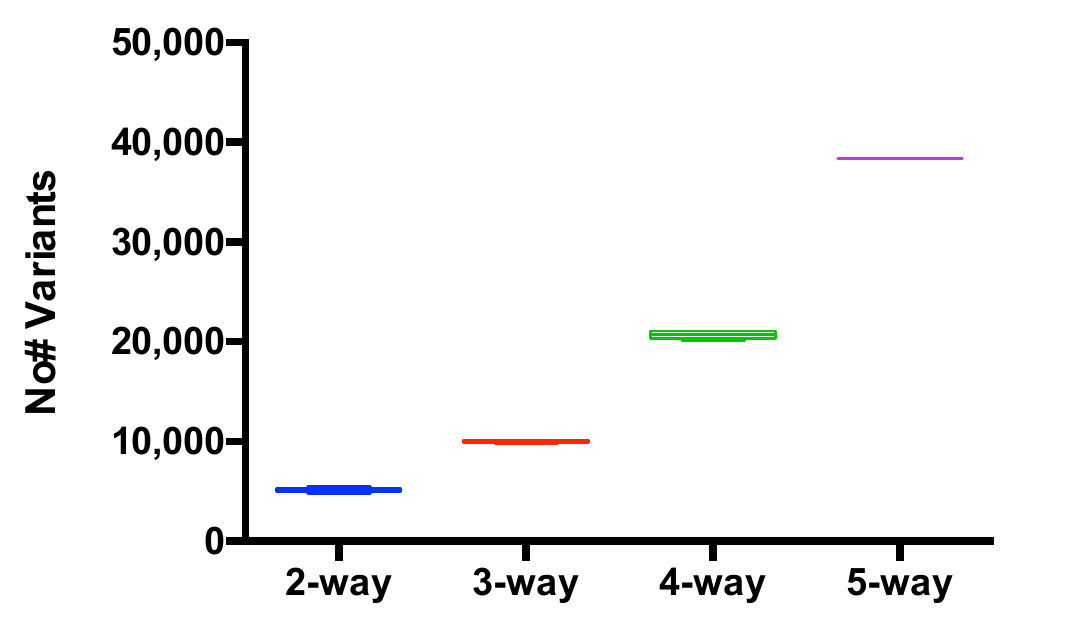}}
                \vspace{-0.5cm}
        \caption{Comparison of the total number of variants generated by avocado during the testing with respect to a $t$-way test suite \centering }
        \label{Figure:VariantBoxPlot}
          \end{subfigure}
             \hfill
    \begin{subfigure}[b]{0.3\textwidth}
    	\fbox{\includegraphics[width=\textwidth]					{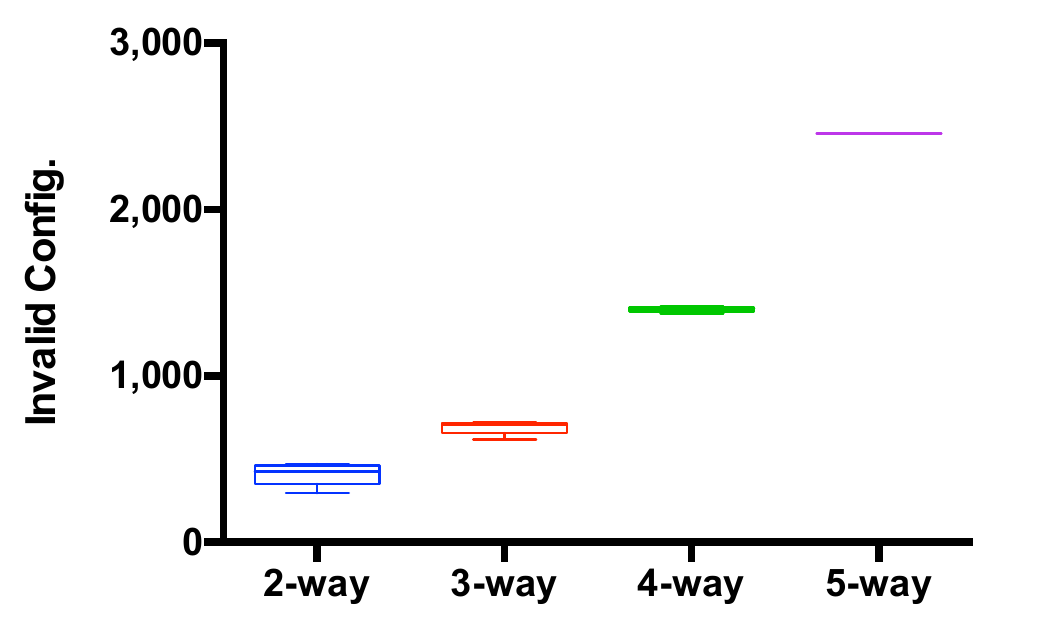}}
                \vspace{-0.5cm}
        \caption{Comparison of the number of invalid configurations detected by Avocado with respect to a $t$-way test suite \centering }
        \label{Figure:FaultBoxPlot}
    \end{subfigure}
\caption{Evaluation results of the CIT plugin}
\label{overallEvaluation}
\end{figure*}

Figure \ref{overallEvaluation} shows the results of the evaluation for criteria 2-4 above. To evaluate the performance of the testing process, we compared the execution time of each t-way test suite. This time represents the total time the Avocado framework takes to generate, execute, and validate the test cases. Figure \ref{Figure:ExecutionTimeBox} shows the box plot analysis to predict the performance and compare the execution time.

The box plot in Figure \ref{Figure:ExecutionTimeBox} reveals a number of salient characteristics of the Avocado and also CIT in general. We can see that the execution time increases with the interaction strength as the size of the test suite is increasing towards the exhaustive (5-way) test suite. Even though the lower quartile of the 3-way test suite is near to the upper quartile of the 2-way test suite,  the time of execution has a notable difference among the test suites. Also, we can see that the deviation and differences among individual test execution time within the same $t$-way test suite are not more significant as the top-to-bottom whisker width is small.

To test the efficiency of the multiplexer integration with test generation, we have compared the number of variants generated for the SUT. Here, the number of the variant is computed by the number of total variants multiplied by the number of test cases generated for a specific $t$-way value. For this case study, we have 192 local variants that each test case must run to validate the configurations. For example, there are 200 test cases for the 5-way test suite. Hence, the total number of variants generated is $200 \times 192 = 38400$ variants to be run for the 5-way testing. Figure \ref{Figure:VariantBoxPlot} shows the box plot analysis for the total number of variants generated for each test suite.

From Figure \ref{Figure:VariantBoxPlot}, we can see a clear difference among the generated variants. There is not even a matching between the lower and upper quartiles of two different variant sets, which in turn shows the test generator and also the multiplexer efficiency in generating variants.

Finally, to evaluate the effectiveness of finding invalid configurations in the case study, we compared the output of Avocado for different test suites. Figure \ref{Figure:FaultBoxPlot} shows a box plot analysis graph to compare different $t$-way test suites. To give a better understanding of the number and proportion of these values, Table \ref{exactNoVariants} shows median values of the variants' number, invalid configurations out of these variants, and the ratio between them. Note that, as mentioned in Section~\ref{subsec:object}, the counts and ratios in Table~\ref{exactNoVariants} do not refer to errors in the code, but to configurations that are not valid in the virtualization environment.

\begin{table}
\centering
\caption{Median values of the generated variants and invalid configurations found with the ration between them}
\small
\begin{tabular}{|c|c|c|c|}
\hline 
Test Suite & \#Variants  & \#Invalid Cofig. & Ratio \%\tabularnewline
\hline 
\hline 
2-way & 5184 & 425 & 8.19\tabularnewline
\hline 
3-way & 9984 & 705 & 7.06\tabularnewline
\hline 
4-way & 20736 & 1405 & 6.77\tabularnewline
\hline 
5-way & 38400 & 2455 & 6.39\tabularnewline
\hline 
\end{tabular}

\label{exactNoVariants}
\end{table}

An important observation can be found in Figure \ref{Figure:FaultBoxPlot} and also Table \ref{exactNoVariants} . Here, we note that the number of detected invalid configurations increases with the growth of the combinatorial interaction strength. We found that the total number of invalid configurations in the SUT is 2455 when the full strength (i.e., 5-way) is considered. As can be predicted in the upper quartile, in the best case, the 2-way test suite can detect 470 invalid configurations, which is almost 19\% of the total faults. In the same way, the 3-way test suite can detect nearly 29\% of total invalid configurations in the best case, while 4-way can detect almost 58\% of total invalid configurations in the best case. We can observe that for this application, the $t$-way test suite can reduce the size and the execution time of the test suites dramatically. However, the full strength combinatorial test suite is necessary to detect all the invalid configurations. Another important observation is that using combinations of input configuration setting values is an effective way to identify invalid configurations. Here, Avocado presents an excellent choice for automating configuration validation and testing. 

We can see from the results that more invalid configurations are found as the test size and combination strength increase. However, it is not immediately clear from these results how many of those configurations detected at 5-way are the result of containing a particular 2-way, 3-way, or 4-way combination that detects an invalid configuration. As it is clear from Figure \ref{CoverageMeasureByNIST}, a configuration with some particular combination strength will also exist in other higher strength combinations of configurations. For example, if we have five binary variables, A, B, C, D, E, and a 2-way combination A=0, B=1 and that results in an invalid configuration, then any 3-way or higher strength combination that includes these values for A and B will also be detected as an invalid configuration. In particular, the 3-way combinations ABC = 010, ABC = 011, ABD = 010, ABD = 011, ABE = 010, and ABE = 011 would all include the invalid combination. Therefore a test such as ABCDE = 01011 would include three 3-way combinations, three 4-way combinations, and one 5-way combination that detect an invalid configuration, but these counts are redundant because only the 2-way combination AB=01 is needed for detection. It is easy to see how these multiple counting situations would increase with a large number of variables. To address this situation and provide a deeper analysis of the test cases, we have reviewed those invalid or failed configurations to determine if this situation is occurring. We have developed a simple open source tool\footnote{https://github.com/bestoun/CombinatorialCounter} that counts the occurrence of the combinations in any $t$-way test suite based on the strength. Table \ref{twayAnalysisTable} shows the result of this analysis.

\begin{table}
\centering
\caption{Invalid Configuration Identification Test Case Results }

\small
\begin{tabular}{|l|c|c|c|c|}
\hline 
 & 2-way & 3-way & 4-way & 5-way\tabularnewline
\hline 
\hline 
Test cases: & 470 & 722 & 1415 & 2455\tabularnewline
\hline 
removing ($t$-1)-way & - & 376 & 960 & 2113\tabularnewline
\hline 
removing ($t$-2)-way & - & - & 572 & 768\tabularnewline
\hline 
removing ($t$-3)-way & - & - & - & 382\tabularnewline
\hline 
\end{tabular}
\label{twayAnalysisTable}
\end{table}

Table \ref{twayAnalysisTable} shows the result of analyzing the test suites. We run the test suites with Avocado and monitor the output of each test case. As mentioned previously, these test suites are used with the multiplexer to produce the variants' set. We identified the invalid configurations in the variant set; then we analyzed each configuration case for covered combinations. As shown in the table, we found and isolated the number of invalid variants' combinations in the configurations. Hence, the numbers used in Table \ref{twayAnalysisTable} are the variants' combinations after multiplication by the configuration test suite.

We have addressed the size of the invalid configurations for each $t$-way test suite. In addition, we have addressed this size of an invalid configuration after removing the repeated lower $t$-way combinations. Here, we used $t-$1, 2, and 3 test suites. For example, for the 5-way test suite, we also generate the 4-way, 3-way, and 2-way combinations and compared them with the used equivalent test suites to identify the repeated test cases based on the tuples.

As we note from the results in Table \ref{twayAnalysisTable}, some configurations are failed (determined to be invalid) in the lower strength of combinations, and they are also repeated in the higher strength test suite. For example, using the analysis tool on the invalid 5-way test cases, we found that 342 repeated 4-way invalid configurations out of those 2455, which results in 2113 5-way configurations after removing them. Also, there are 1345 repeated 3-way invalid configurations out of those 2113 remaining configurations, which results in 768 5-way configurations after removing them. In the same way, there are 368 repeated 2-way invalid configurations out of those 768 remaining configurations, which results in 382 5-way configurations after removing them.  

We can conclude from Table \ref{twayAnalysisTable} that for this application, lower strength combinations are responsible for part of those invalid configurations; however, higher interaction strength (greater value of $t$) combinations are needed to detect all the possible configuration failures. In fact, this shows that unlike studies of fault detection in the literature, e.g. \cite{kuhn2004software}, running only pairwise (2-way) test cases is not enough for this application to trigger most of the failures. The reason for this difference is that other applications of combinatorial testing have generally been for detecting errors in code.  In this case, however, testing addressed detection of configurations that could not be supported in the virtual machine environment, rather than detecting code flaws.  This is a different use of combinatorial methods, but $t$-way testing was shown to be highly effective.  Like any other configurable system, for larger configurations of this application in the industry, running full exhaustive testing in most cases is impossible. Hence, running lower combination strength test suites with Avocado is an option to assure quality and avoid triggering configuration failure.

\section{Concluding Remarks}\label{Conclusion}

We have demonstrated a method of automating the combinatorial interaction testing process, using the open source Avocado testing framework with CIT capabilities implemented in a plugin. Within Avocado CIT, the tester needs only to establish the environment of the application to be tested. The Avocado framework was used for validating virtual machine configurations for Qemu, demonstrating that Avocado can be a cost-effective tool for automating this essential step in the virtualizer setup.

Avocado is a flexible and customizable framework in which other capabilities, features, algorithms,  and tools can be added easily through a plugin. We plan to add constraint handling capabilities to the framework through a constraint solver. Avocado is a freely available open source project freely available\footnote{https://github.com/avocado-framework/avocado}.

\begin{acks}
This research is funded by Red Hat Czech s.r.o. as a collaboration project with Software Testing Intelligent Lab (STILL) in CVUT and part of Avocado testing framework project. Products may be identified in this document, but identification does not imply recommendation or endorsement by NIST, nor that the products identified are necessarily the best available for the purpose.
\end{acks}

%
\bibliographystyle{ACM-Reference-Format}
\bibliography{sample-base}


\begin{thebibliography}{23}


\ifx \showCODEN    \undefined \def \showCODEN     #1{\unskip}     \fi
\ifx \showDOI      \undefined \def \showDOI       #1{#1}\fi
\ifx \showISBNx    \undefined \def \showISBNx     #1{\unskip}     \fi
\ifx \showISBNxiii \undefined \def \showISBNxiii  #1{\unskip}     \fi
\ifx \showISSN     \undefined \def \showISSN      #1{\unskip}     \fi
\ifx \showLCCN     \undefined \def \showLCCN      #1{\unskip}     \fi
\ifx \shownote     \undefined \def \shownote      #1{#1}          \fi
\ifx \showarticletitle \undefined \def \showarticletitle #1{#1}   \fi
\ifx \showURL      \undefined \def \showURL       {\relax}        \fi
\providecommand\bibfield[2]{#2}
\providecommand\bibinfo[2]{#2}
\providecommand\natexlab[1]{#1}
\providecommand\showeprint[2][]{arXiv:#2}

\bibitem[\protect\citeauthoryear{Ahmed and Zamli}{Ahmed and Zamli}{2010}]%
        {Ahmed:2010:PTS}
\bibfield{author}{\bibinfo{person}{Bestoun~S. Ahmed} {and}
  \bibinfo{person}{Kamal~Z. Zamli}.} \bibinfo{year}{2010}\natexlab{}.
\newblock \showarticletitle{PSTG: A T-Way Strategy Adopting Particle Swarm
  Optimization}. In \bibinfo{booktitle}{\emph{Proceedings of the 2010 Fourth
  Asia International Conference on Mathematical/Analytical Modelling and
  Computer Simulation}} \emph{(\bibinfo{series}{AMS '10})}.
  \bibinfo{publisher}{IEEE Computer Society}, \bibinfo{address}{Washington, DC,
  USA}, \bibinfo{pages}{1--5}.
\newblock
\showISBNx{978-0-7695-4062-7}
\urldef\tempurl%
\url{https://doi.org/10.1109/AMS.2010.14}
\showDOI{\tempurl}


\bibitem[\protect\citeauthoryear{Ahmed and Zamli}{Ahmed and Zamli}{2011}]%
        {Ahmed:2011}
\bibfield{author}{\bibinfo{person}{Bestoun~S. Ahmed} {and}
  \bibinfo{person}{Kamal~Z. Zamli}.} \bibinfo{year}{2011}\natexlab{}.
\newblock \showarticletitle{A Variable Strength Interaction Test Suites
  Generation Strategy Using Particle Swarm Optimization}.
\newblock \bibinfo{journal}{\emph{J. Syst. Softw.}} \bibinfo{volume}{84},
  \bibinfo{number}{12} (\bibinfo{date}{Dec.} \bibinfo{year}{2011}),
  \bibinfo{pages}{2171--2185}.
\newblock
\showISSN{0164-1212}
\urldef\tempurl%
\url{https://doi.org/10.1016/j.jss.2011.06.004}
\showDOI{\tempurl}


\bibitem[\protect\citeauthoryear{Ahmed, Zamli, Afzal, and Bures}{Ahmed
  et~al\mbox{.}}{2017}]%
        {BestouIEEAcess}
\bibfield{author}{\bibinfo{person}{Bestoun~S. Ahmed}, \bibinfo{person}{Kamal~Z.
  Zamli}, \bibinfo{person}{Wasif Afzal}, {and} \bibinfo{person}{Miroslav
  Bures}.} \bibinfo{year}{2017}\natexlab{}.
\newblock \showarticletitle{Constrained Interaction Testing: A Systematic
  Literature Study}.
\newblock \bibinfo{journal}{\emph{IEEE Access}}  \bibinfo{volume}{5}
  (\bibinfo{year}{2017}), \bibinfo{pages}{25706--25730}.
\newblock
\urldef\tempurl%
\url{https://doi.org/10.1109/ACCESS.2017.2771562}
\showDOI{\tempurl}


\bibitem[\protect\citeauthoryear{Ahmed, Zamli, and Lim}{Ahmed
  et~al\mbox{.}}{2012}]%
        {Ahmed:2012}
\bibfield{author}{\bibinfo{person}{Bestoun~S. Ahmed}, \bibinfo{person}{Kamal~Z.
  Zamli}, {and} \bibinfo{person}{Chee~Peng Lim}.}
  \bibinfo{year}{2012}\natexlab{}.
\newblock \showarticletitle{Application of Particle Swarm Optimization to
  Uniform and Variable Strength Covering Array Construction}.
\newblock \bibinfo{journal}{\emph{Applied Soft Computing}}
  \bibinfo{volume}{12}, \bibinfo{number}{4} (\bibinfo{date}{April}
  \bibinfo{year}{2012}), \bibinfo{pages}{1330--1347}.
\newblock
\showISSN{1568-4946}
\urldef\tempurl%
\url{https://doi.org/10.1016/j.asoc.2011.11.029}
\showDOI{\tempurl}


\bibitem[\protect\citeauthoryear{Bansal, Mittal, Sabharwal, and Koul}{Bansal
  et~al\mbox{.}}{2014}]%
        {Bansal2014}
\bibfield{author}{\bibinfo{person}{P. Bansal}, \bibinfo{person}{N. Mittal},
  \bibinfo{person}{A. Sabharwal}, {and} \bibinfo{person}{S. Koul}.}
  \bibinfo{year}{2014}\natexlab{}.
\newblock \showarticletitle{Integrating greedy based approach with genetic
  algorithm to generate mixed covering arrays for pair-wise testing}. In
  \bibinfo{booktitle}{\emph{2014 Seventh International Conference on
  Contemporary Computing (IC3)}}. \bibinfo{pages}{629--634}.
\newblock
\urldef\tempurl%
\url{https://doi.org/10.1109/IC3.2014.6897246}
\showDOI{\tempurl}


\bibitem[\protect\citeauthoryear{Bartholomew}{Bartholomew}{2013}]%
        {bartholomew2013industry}
\bibfield{author}{\bibinfo{person}{Redge Bartholomew}.}
  \bibinfo{year}{2013}\natexlab{}.
\newblock \showarticletitle{An industry proof-of-concept demonstration of
  automated combinatorial test}. In \bibinfo{booktitle}{\emph{Proceedings of
  the 8th International Workshop on Automation of Software Test}}. IEEE Press,
  \bibinfo{pages}{118--124}.
\newblock


\bibitem[\protect\citeauthoryear{Borodai and Grunskii}{Borodai and
  Grunskii}{1992}]%
        {Borodai1992}
\bibfield{author}{\bibinfo{person}{S.~Yu. Borodai} {and} \bibinfo{person}{I.~S.
  Grunskii}.} \bibinfo{year}{1992}\natexlab{}.
\newblock \showarticletitle{Recursive generation of locally complete tests}.
\newblock \bibinfo{journal}{\emph{Cybernetics and Systems Analysis}}
  \bibinfo{volume}{28}, \bibinfo{number}{4} (\bibinfo{date}{01 Jul}
  \bibinfo{year}{1992}), \bibinfo{pages}{504--508}.
\newblock
\showISSN{1573-8337}
\urldef\tempurl%
\url{https://doi.org/10.1007/BF01124983}
\showDOI{\tempurl}


\bibitem[\protect\citeauthoryear{Bures and Ahmed}{Bures and Ahmed}{2017}]%
        {Bures2017}
\bibfield{author}{\bibinfo{person}{Miroslav Bures} {and}
  \bibinfo{person}{Bestoun~S. Ahmed}.} \bibinfo{year}{2017}\natexlab{}.
\newblock \showarticletitle{On the Effectiveness of Combinatorial Interaction
  Testing: A Case Study}. In \bibinfo{booktitle}{\emph{2017 IEEE International
  Conference on Software Quality, Reliability and Security Companion (QRS-C)}}.
  \bibinfo{pages}{69--76}.
\newblock
\urldef\tempurl%
\url{https://doi.org/10.1109/QRS-C.2017.20}
\showDOI{\tempurl}


\bibitem[\protect\citeauthoryear{Forbes, Lawrence, Lei, Kacker, and
  Kuhn}{Forbes et~al\mbox{.}}{2008}]%
        {Forbes2008287}
\bibfield{author}{\bibinfo{person}{M. Forbes}, \bibinfo{person}{J. Lawrence},
  \bibinfo{person}{Y. Lei}, \bibinfo{person}{R.N. Kacker}, {and}
  \bibinfo{person}{D.R. Kuhn}.} \bibinfo{year}{2008}\natexlab{}.
\newblock \showarticletitle{Refining the in-parameter-order strategy for
  constructing covering arrays}.
\newblock \bibinfo{journal}{\emph{Journal of Research of the National Institute
  of Standards and Technology}} \bibinfo{volume}{113}, \bibinfo{number}{5}
  (\bibinfo{year}{2008}), \bibinfo{pages}{287--297}.
\newblock
\newblock
\shownote{cited By 91.}


\bibitem[\protect\citeauthoryear{Gargantini and Vavassori}{Gargantini and
  Vavassori}{2012}]%
        {gargantini2012citlab}
\bibfield{author}{\bibinfo{person}{Angelo Gargantini} {and}
  \bibinfo{person}{Paolo Vavassori}.} \bibinfo{year}{2012}\natexlab{}.
\newblock \showarticletitle{CitLab: a laboratory for combinatorial interaction
  testing}. In \bibinfo{booktitle}{\emph{Software Testing, Verification and
  Validation (ICST), 2012 IEEE Fifth International Conference on}}. IEEE,
  \bibinfo{pages}{559--568}.
\newblock


\bibitem[\protect\citeauthoryear{Hartman and Raskin}{Hartman and
  Raskin}{2004}]%
        {HARTMAN2004149}
\bibfield{author}{\bibinfo{person}{Alan Hartman} {and} \bibinfo{person}{Leonid
  Raskin}.} \bibinfo{year}{2004}\natexlab{}.
\newblock \showarticletitle{Problems and algorithms for covering arrays}.
\newblock \bibinfo{journal}{\emph{Discrete Mathematics}} \bibinfo{volume}{284},
  \bibinfo{number}{1} (\bibinfo{year}{2004}), \bibinfo{pages}{149 -- 156}.
\newblock
\showISSN{0012-365X}
\urldef\tempurl%
\url{https://doi.org/10.1016/j.disc.2003.11.029}
\showDOI{\tempurl}
\newblock
\shownote{Special Issue in Honour of Curt Lindner on His 65th Birthday.}


\bibitem[\protect\citeauthoryear{Huang, Xie, Chen, and Lu}{Huang
  et~al\mbox{.}}{2012}]%
        {Huang2012}
\bibfield{author}{\bibinfo{person}{R. Huang}, \bibinfo{person}{X. Xie},
  \bibinfo{person}{T.~Y. Chen}, {and} \bibinfo{person}{Y. Lu}.}
  \bibinfo{year}{2012}\natexlab{}.
\newblock \showarticletitle{Adaptive Random Test Case Generation for
  Combinatorial Testing}. In \bibinfo{booktitle}{\emph{2012 IEEE 36th Annual
  Computer Software and Applications Conference}}. \bibinfo{pages}{52--61}.
\newblock
\showISSN{0730-3157}
\urldef\tempurl%
\url{https://doi.org/10.1109/COMPSAC.2012.15}
\showDOI{\tempurl}


\bibitem[\protect\citeauthoryear{Huller}{Huller}{2000}]%
        {Huller00reducingtime}
\bibfield{author}{\bibinfo{person}{Jerry Huller}.}
  \bibinfo{year}{2000}\natexlab{}.
\newblock \showarticletitle{Reducing time to market with combinatorial design
  method testing}. In \bibinfo{booktitle}{\emph{In Proceedings of the 2000
  International Council on Systems Engineering (INCOSE) Conference}}.
  \bibinfo{pages}{16--20}.
\newblock


\bibitem[\protect\citeauthoryear{Kuhn, Wallace, and Gallo}{Kuhn
  et~al\mbox{.}}{2004}]%
        {kuhn2004software}
\bibfield{author}{\bibinfo{person}{D~Richard Kuhn}, \bibinfo{person}{Dolores~R
  Wallace}, {and} \bibinfo{person}{Albert~M Gallo}.}
  \bibinfo{year}{2004}\natexlab{}.
\newblock \showarticletitle{Software fault interactions and implications for
  software testing}.
\newblock \bibinfo{journal}{\emph{IEEE transactions on software engineering}}
  \bibinfo{volume}{30}, \bibinfo{number}{6} (\bibinfo{year}{2004}),
  \bibinfo{pages}{418--421}.
\newblock


\bibitem[\protect\citeauthoryear{Lei, Kacker, Kuhn, Okun, and Lawrence}{Lei
  et~al\mbox{.}}{2008}]%
        {Lei:2008:IET}
\bibfield{author}{\bibinfo{person}{Yu Lei}, \bibinfo{person}{Raghu Kacker},
  \bibinfo{person}{D.~Richard Kuhn}, \bibinfo{person}{Vadim Okun}, {and}
  \bibinfo{person}{James Lawrence}.} \bibinfo{year}{2008}\natexlab{}.
\newblock \showarticletitle{IPOG-IPOG-D: Efficient Test Generation for
  Multi-way Combinatorial Testing}.
\newblock \bibinfo{journal}{\emph{Softw. Test. Verif. Reliab.}}
  \bibinfo{volume}{18}, \bibinfo{number}{3} (\bibinfo{date}{Sept.}
  \bibinfo{year}{2008}), \bibinfo{pages}{125--148}.
\newblock
\showISSN{0960-0833}
\urldef\tempurl%
\url{https://doi.org/10.1002/stvr.v18:3}
\showDOI{\tempurl}


\bibitem[\protect\citeauthoryear{Li, Gao, Wong, Yang, and Li}{Li
  et~al\mbox{.}}{2016}]%
        {li2016applying}
\bibfield{author}{\bibinfo{person}{Xuelin Li}, \bibinfo{person}{Ruizhi Gao},
  \bibinfo{person}{W~Eric Wong}, \bibinfo{person}{Chunhui Yang}, {and}
  \bibinfo{person}{Dong Li}.} \bibinfo{year}{2016}\natexlab{}.
\newblock \showarticletitle{Applying combinatorial testing in industrial
  settings}. In \bibinfo{booktitle}{\emph{Software Quality, Reliability and
  Security (QRS), 2016 IEEE International Conference on}}. IEEE,
  \bibinfo{pages}{53--60}.
\newblock


\bibitem[\protect\citeauthoryear{Nurmela}{Nurmela}{2004}]%
        {NURMELA2004143}
\bibfield{author}{\bibinfo{person}{Kari~J. Nurmela}.}
  \bibinfo{year}{2004}\natexlab{}.
\newblock \showarticletitle{Upper bounds for covering arrays by tabu search}.
\newblock \bibinfo{journal}{\emph{Discrete Applied Mathematics}}
  \bibinfo{volume}{138}, \bibinfo{number}{1} (\bibinfo{year}{2004}),
  \bibinfo{pages}{143 -- 152}.
\newblock
\showISSN{0166-218X}
\urldef\tempurl%
\url{https://doi.org/10.1016/S0166-218X(03)00291-9}
\showDOI{\tempurl}
\newblock
\shownote{Optimal Discrete Structures and Algorithms.}


\bibitem[\protect\citeauthoryear{Schubert}{Schubert}{2004}]%
        {Schubert2004}
\bibfield{author}{\bibinfo{person}{Ulrich~S. Schubert}.}
  \bibinfo{year}{2004}\natexlab{}.
\newblock \showarticletitle{Experimental Design for Combinatorial and High
  Throughput Materials Development. Edited by James N. Cawse.}
\newblock \bibinfo{journal}{\emph{Angewandte Chemie International Edition}}
  \bibinfo{volume}{43}, \bibinfo{number}{32} (\bibinfo{year}{2004}),
  \bibinfo{pages}{4123--4123}.
\newblock
\urldef\tempurl%
\url{https://doi.org/10.1002/anie.200385086}
\showDOI{\tempurl}


\bibitem[\protect\citeauthoryear{Sherif}{Sherif}{2016}]%
        {Anwar2016}
\bibfield{author}{\bibinfo{person}{Anwar Sherif}.}
  \bibinfo{year}{2016}\natexlab{}.
\newblock \showarticletitle{Combinatorial Testing: Implementations in Solutions
  Testing}. In \bibinfo{booktitle}{\emph{2016 IEEE Ninth International
  Conference on Software Testing, Verification and Validation Workshops
  (ICSTW)}}. \bibinfo{pages}{59--64}.
\newblock
\urldef\tempurl%
\url{https://doi.org/10.1109/ICSTW.2016.39}
\showDOI{\tempurl}


\bibitem[\protect\citeauthoryear{Shiba, Tsuchiya, and Kikuno}{Shiba
  et~al\mbox{.}}{2004}]%
        {Shiba:2004}
\bibfield{author}{\bibinfo{person}{Toshiaki Shiba}, \bibinfo{person}{Tatsuhiro
  Tsuchiya}, {and} \bibinfo{person}{Tohru Kikuno}.}
  \bibinfo{year}{2004}\natexlab{}.
\newblock \showarticletitle{Using Artificial Life Techniques to Generate Test
  Cases for Combinatorial Testing}. In \bibinfo{booktitle}{\emph{Proceedings of
  the 28th Annual International Computer Software and Applications Conference -
  Volume 01}} \emph{(\bibinfo{series}{COMPSAC '04})}. \bibinfo{publisher}{IEEE
  Computer Society}, \bibinfo{address}{Washington, DC, USA},
  \bibinfo{pages}{72--77}.
\newblock
\showISBNx{0-7695-2209-2-1}


\bibitem[\protect\citeauthoryear{Sulaiman and Ahmed}{Sulaiman and
  Ahmed}{2013}]%
        {Sulaiman2013}
\bibfield{author}{\bibinfo{person}{Diary~R. Sulaiman} {and}
  \bibinfo{person}{Bestoun~S. Ahmed}.} \bibinfo{year}{2013}\natexlab{}.
\newblock \showarticletitle{Using the combinatorial optimization approach for
  DVS in high performance processors}. In \bibinfo{booktitle}{\emph{2013 The
  International Conference on Technological Advances in Electrical, Electronics
  and Computer Engineering (TAEECE)}}. \bibinfo{pages}{105--109}.
\newblock
\urldef\tempurl%
\url{https://doi.org/10.1109/TAEECE.2013.6557204}
\showDOI{\tempurl}


\bibitem[\protect\citeauthoryear{Yilmaz, Cohen, and Porter}{Yilmaz
  et~al\mbox{.}}{2004}]%
        {Yilmaz:2004}
\bibfield{author}{\bibinfo{person}{Cemal Yilmaz}, \bibinfo{person}{Myra~B.
  Cohen}, {and} \bibinfo{person}{Adam Porter}.}
  \bibinfo{year}{2004}\natexlab{}.
\newblock \showarticletitle{Covering Arrays for Efficient Fault
  Characterization in Complex Configuration Spaces}.
\newblock \bibinfo{journal}{\emph{SIGSOFT Softw. Eng. Notes}}
  \bibinfo{volume}{29}, \bibinfo{number}{4} (\bibinfo{date}{July}
  \bibinfo{year}{2004}), \bibinfo{pages}{45--54}.
\newblock
\showISSN{0163-5948}
\urldef\tempurl%
\url{https://doi.org/10.1145/1013886.1007519}
\showDOI{\tempurl}


\bibitem[\protect\citeauthoryear{Yuan, Cohen, and Memon}{Yuan
  et~al\mbox{.}}{2011}]%
        {Yuan:2011}
\bibfield{author}{\bibinfo{person}{Xun Yuan}, \bibinfo{person}{Myra~B. Cohen},
  {and} \bibinfo{person}{Atif~M. Memon}.} \bibinfo{year}{2011}\natexlab{}.
\newblock \showarticletitle{GUI Interaction Testing: Incorporating Event
  Context}.
\newblock \bibinfo{journal}{\emph{IEEE Transactions on Software Engineering}}
  \bibinfo{volume}{37}, \bibinfo{number}{4} (\bibinfo{date}{July}
  \bibinfo{year}{2011}), \bibinfo{pages}{559--574}.
\newblock
\showISSN{0098-5589}
\urldef\tempurl%
\url{https://doi.org/10.1109/TSE.2010.50}
\showDOI{\tempurl}


\end{thebibliography}

\end{document}